\documentclass{article}
\usepackage{amsmath}

\begin{document}

\title{Fractional Hadamard transform with continuous variables in the
context of quantum optics\thanks{{\small Project supported by the National
Natural Science Foundation of China (Grant Nos 10775097 and 10874174) and
the Research Foundation of the Education Department of Jiangxi Province of
China.}}}
\author{Li-yun Hu$\thanks{{\small Corresponding author. E-mail:
hlyun2008@126.com; hlyun@jxnu.edu.cn. }}$, Xue-xiang Xu, and Shan-jun Ma \\
%EndAName
{\small College of Physics and Communication Electronics, Jiangxi Normal
University, Nanchang 330022, China}}
\maketitle

\begin{abstract}
{\small We introduce the quantum fractional Hadamard transform with
continuous variables. It is found that the corresponding quantum fractional
Hadamard operator can be decomposed into a single-mode fractional operator
and two single-mode squeezing operators. This is extended to the entangled
case by using the bipartite entangled state representation. The new
transformation presents more flexibility to represent signals in the
fractional Hadamard domain with extra freedom provided by an angle and
two-squeezing parameters.}
\end{abstract}

Keywords: fractional Hadamard transform, fractional Hadamard operator, the
additivity of operator

PACC: 0367, 4250

\section{Introduction}

Fractional Fourier transform (FrFT) is a generalization of the ordinary
Fourier transform, which has been used in signal processing and image
manipulations \cite{1,2}. The concept of the FrFT was originally described
by Condon \cite{3} and was later introduced for signal processing by Namias
\cite{4} as a Fourier transform of fractional order. The 1-dimension FrFT of
$\alpha $-order is defined in Refs.\cite{5,6} as
\begin{equation}
g\left( x^{\prime }\right) =\sqrt{\frac{e^{\mathtt{i}\left( \frac{\pi }{2}%
-\alpha \right) }}{2\pi \sin \alpha }}\int e^{-\frac{\mathtt{i}\left(
x^{\prime 2}+x^{2}\right) }{2\tan \alpha }+\frac{\mathtt{i}xx^{\prime }}{%
\sin \alpha }}f\left( x\right) \mathtt{d}x.  \label{f1}
\end{equation}%
The usual Fourier transform is a special case with order $\alpha =\pi /2$.
On the other hand, many orthogonal transform have been successfully used in
signal processing, such as discrete cosine transform \cite{7}, discrete
Hartley transform \cite{8} and Hadamard transform.

Hadamard transform is not only an important tool in classical signal
processing, but also is of great importance for quantum computation
applications \cite{9}. This transform, used to go from the position basis $%
\left\vert x\right\rangle $ to the momentum basis, is defined as \cite{10,11}%
\begin{equation}
\mathcal{F}\left\vert x\right\rangle =\frac{1}{\sqrt{\pi }\sigma }%
\int_{-\infty }^{\infty }e^{2\mathtt{i}xy/\sigma ^{2}}\left\vert
y\right\rangle \mathtt{d}y,  \label{f2}
\end{equation}%
where $\sigma $ is the scale length (also makes the expression in the
exponential dimensionless), $\left\vert x\right\rangle $ and $\left\vert
y\right\rangle $ are the eigenvector of coordinate operator $X.$ In Ref.\cite%
{12}, the explicit form of $\mathcal{F}$ has been derived by using the
technique of integration within an ordered product (IWOP) of operators \cite%
{13,14,15}, and it is found that it can be decomposed into a single-mode
squeezing operator and a position-momentum mutual transform operator, i.e., $%
\mathcal{F=}S_{1}^{-1}(-1)^{i\pi a^{\dag }a/2}$. In addition, the two-mode
Hadamard transform with continuous variables is also introduced by using the
bipartite entangled state representation, whose Hadamard operator involves a
two-mode squeezing operator and a mutual transform operator.

In this paper, we shall introduce the continuous fractional Hadamard
transform (CFrHT), which is a generalization of the usual Hadamard transform
in Eq.(\ref{f2}). The development of the CFrHT is based upon the same spirit
of continuous fractional Fourier transform (CFrFT). Then the CFrHT operator
(CFrHTO) is derived by using the IWOP technique, and its properties are
analyzed. It is found that the CFrHTO can be decomposed into a single-mode
fractional operator $e^{\mathtt{i}\alpha a^{\dagger }a}$ and two single-mode
squeezing operators. On the other hand, since the publication of the paper
of Einstein, Podolsky and Rosen (EPR) in 1935 \cite{15b}, arguing the
incompleteness of quantum mechanics, the conception of entanglement has
become more and more fascinating and important as it plays a central role in
quantum imformation and quantum computation, we also shall introduce the
two-mode CFrHO in bipartite entangled state representation, which turns out
to involve the two fractional operators and two two-mode squeezing operators.

Our work is arranged as follows. In section 2, for the single-mode case, the
normally ordered fractional Hadamard operator is derived by using the IWOP
technique. The properties of fractional Hadamard operator is discussed in
section 3, such as the unitarity, the decomposition of the CFrHO and its
transform relation. Then the single-mode case is extended to two-mode case
in section 4 and some similar discussions to singled-mode case are
presented. Section 5 is devoted to exploring the measurements for the output
states from the CFrHT. Conclusions are involved in the last section.

\section{Normally Ordered Fractional Hadamard Operator}

In this section, we first introduce the continuous fractional Hadamard
transform (CFrHT), i.e.,

\begin{equation}
\mathcal{H}_{\alpha }\left( \mu ,\nu \right) \left\vert x\right\rangle =%
\sqrt{\frac{e^{\mathtt{i}\left( \frac{\pi }{2}-\alpha \right) }}{2\pi \mu
\nu \sin \alpha }}\int_{-\infty }^{\infty }\exp \left\{ -\frac{\mathtt{i}%
\left( x^{2}/\mu ^{2}+y^{2}/\nu ^{2}\right) }{2\tan \alpha }+\frac{\mathtt{i}%
xy}{\mu \nu \sin \alpha }\right\} \left\vert y\right\rangle \mathtt{d}y,
\label{f3}
\end{equation}%
where $\mu ,\nu $ are the scale length (also make the expression in the
exponential dimensionless), $\alpha $ is an angle, and $\mathcal{H}^{\alpha
}\left( \mu ,\nu \right) $ is called the CFrHT operator (CFrHO). In
particular, when $\alpha =\pi /2$ and $\mu =\nu =\sigma /\sqrt{2},$ Eq.(\ref%
{f3}) just reduces to Eq.(\ref{f2}).

In order to find the explicit expression of the CFrHO, multiplying Eq.(\ref%
{f3}) by the bra $\int \mathtt{d}x\left\langle x\right\vert $ from the
rights in two-side, where $\left\vert y\right\rangle $ and $\left\vert
x\right\rangle $ are coordinate eigenvectors, $X\left\vert x\right\rangle
=x\left\vert x\right\rangle $, and
\begin{equation}
\left\vert x\right\rangle =\pi ^{-1/4}\exp \left\{ -\frac{x^{2}}{2}+\sqrt{2}%
xa^{\dagger }-\frac{a^{\dagger 2}}{2}\right\} \left\vert 0\right\rangle ,
\label{f4}
\end{equation}%
we can recast the CFrHO $\mathcal{H}_{\alpha }\left( \mu ,\nu \right) $ into
the following integral form,%
\begin{equation}
\mathcal{H}_{\alpha }\left( \mu ,\nu \right) =\sqrt{\frac{e^{\mathtt{i}%
\left( \frac{\pi }{2}-\alpha \right) }}{2\pi \mu \nu \sin \alpha }}%
\int_{-\infty }^{\infty }\exp \left\{ -\frac{\mathtt{i}\left( x^{2}/\mu
^{2}+y^{2}/\nu ^{2}\right) }{2\tan \alpha }+\frac{\mathtt{i}xy}{\mu \nu \sin
\alpha }\right\} \left\vert y\right\rangle \left\langle x\right\vert \mathtt{%
d}x\mathtt{d}y.  \label{f5}
\end{equation}%
Then using the vacuum projector's normal ordering form $\left\vert
0\right\rangle \left\langle 0\right\vert =\colon e^{-a^{\dag }a}\colon $%
(where the symbol $\colon $ $\colon $ denotes the normally ordering) and the
IWOP technique to directly perform the integration, we finally obtain%
\begin{eqnarray}
\mathcal{H}_{\alpha }\left( \mu ,\nu \right) &=&\frac{1}{\pi }\sqrt{\frac{e^{%
\mathtt{i}\left( \frac{\pi }{2}-\alpha \right) }}{2\mu \nu \sin \alpha }}%
\int_{-\infty }^{\infty }\colon \exp \left\{ -\frac{A}{2\mu ^{2}}x^{2}+\sqrt{%
2}xa+\frac{\mathtt{i}xy}{\mu \nu \sin \alpha }\right.  \notag \\
&&-\left. \frac{B}{2\nu ^{2}}y^{2}+\sqrt{2}ya^{\dagger }-\frac{\left(
a^{\dag }+a\right) ^{2}}{2}\right\} \colon \mathtt{d}x\mathtt{d}y  \notag \\
&=&\sqrt{\frac{2\mu \nu e^{\mathtt{i}\left( \frac{\pi }{2}-\alpha \right) }}{%
u\sin \alpha }}\exp \left\{ \left( \frac{\nu ^{2}A}{u}-\frac{1}{2}\right)
a^{\dag 2}\right\}  \notag \\
&&\times \exp \left\{ a^{\dagger }a\ln \frac{\mathtt{i}2\mu \nu }{u\sin
\alpha }\right\} \exp \left\{ \left( \frac{\mu ^{2}B}{u}-\frac{1}{2}\right)
a^{2}\right\} ,  \label{f6}
\end{eqnarray}%
where we have set $\allowbreak A=\mathtt{i}\cot \alpha +\mu ^{2},B=\mathtt{i}%
\cot \alpha +\nu ^{2},u=\csc ^{2}\alpha +AB,$and used the operator identity
in the last step of Eq.(\ref{f6}),

\begin{equation}
\exp \left\{ fa^{\dagger }a\right\} =\colon \exp \left\{ \left(
e^{f}-1\right) a^{\dagger }a\right\} \colon ,  \label{f8}
\end{equation}%
Eq.(\ref{f6}) is the normally ordered form of the CFrHO. In particular, when
$\alpha =\pi /2$ and $\mu =\nu =\sigma /\sqrt{2},$ leading to $A=B=\sigma
^{2}/2,u=1+\sigma ^{4}/4,$ then Eq.(\ref{f6}) becomes%
\begin{eqnarray}
\mathcal{H}_{\pi /2}\left( \sigma /\sqrt{2},\sigma /\sqrt{2}\right) &=&\frac{%
2\sigma }{\sqrt{\sigma ^{4}+4}}\exp \left\{ \frac{\sigma ^{4}-4}{\sigma
^{4}+4}\frac{a^{\dagger 2}}{2}\right\}  \notag \\
&&\times \exp \left\{ a^{\dagger }a\ln \frac{4\mathtt{i}\sigma ^{2}}{\sigma
^{4}+4}\right\} \exp \left\{ \frac{\sigma ^{4}-4}{\sigma ^{4}+4}\frac{a^{2}}{%
2}\right\} ,  \label{f9}
\end{eqnarray}%
which is just the result Eq.(7) in Ref.\cite{12}.

\section{Properties of Fractional Hadamard Operator}

From Eq.(\ref{f5}) one can see that the CFrHO is a unitary one, i.e., $%
\mathcal{H}_{\alpha }\left( \mu ,\nu \right) \mathcal{H}_{\alpha }^{\dagger
}\left( \mu ,\nu \right) =\mathcal{H}_{\alpha }^{\dagger }\left( \mu ,\nu
\right) \mathcal{H}_{\alpha }\left( \mu ,\nu \right) =1.$ In fact, uisng Eq.(%
\ref{f5}) and the orthogonality of coordinate state, $\left\langle x^{\prime
}\right. \left\vert x\right\rangle =\delta \left( x-x^{\prime }\right) $, we
have%
\begin{eqnarray}
\mathcal{H}_{\alpha }\left( \mu ,\nu \right) \mathcal{H}_{\alpha }^{\dagger
}\left( \mu ,\nu \right) &=&\frac{1}{2\pi \mu \nu \sin \alpha }\int_{-\infty
}^{\infty }\exp \left\{ \allowbreak \frac{\mathtt{i}\left( y^{\prime
2}-y^{2}\right) }{2\nu ^{2}\tan \alpha }+\mathtt{i}x\frac{y-y^{\prime }}{\mu
\nu \sin \alpha }\right\} \left\vert y\right\rangle \left\langle y^{\prime
}\right\vert \mathtt{d}x\mathtt{d}y\mathtt{d}y^{\prime }  \notag \\
&=&\frac{1}{\mu \nu \sin \alpha }\int_{-\infty }^{\infty }\delta \left(
\frac{y-y^{\prime }}{\mu \nu \sin \alpha }\right) \exp \left\{ \allowbreak
\frac{\mathtt{i}\left( y^{\prime 2}-y^{2}\right) }{2\nu ^{2}\tan \alpha }%
\right\} \left\vert y\right\rangle \left\langle y^{\prime }\right\vert
\mathtt{d}y\mathtt{d}y^{\prime }  \notag \\
&=&\int_{-\infty }^{\infty }\left\vert y\right\rangle \left\langle
y\right\vert \mathtt{d}y=\mathcal{H}_{\alpha }^{\dagger }\left( \mu ,\nu
\right) \mathcal{H}_{\alpha }\left( \mu ,\nu \right) =1.  \label{f10}
\end{eqnarray}

In order to see clearly its transform relation under the CFrHO, next we
examine its decomposition. Performing the change of variables, $x/\mu
\rightarrow x,$ $y/\nu \rightarrow y$, we can be recast Eq.(\ref{f5}) into
the following form,

\begin{equation}
\mathcal{H}_{\alpha }\left( \mu ,\nu \right) =\sqrt{\frac{\mu \nu e^{\mathtt{%
i}\left( \frac{\pi }{2}-\alpha \right) }}{2\pi \sin \alpha }}\int_{-\infty
}^{\infty }\exp \left\{ -\frac{\mathtt{i}\left( x^{2}+y^{2}\right) }{2\tan
\alpha }+\frac{\mathtt{i}xy}{\sin \alpha }\right\} \left\vert \nu
y\right\rangle \left\langle \mu x\right\vert \mathtt{d}x\mathtt{d}y.
\label{f11}
\end{equation}%
By noticing that the single-mode squeezing operator $S_{1}$ \cite{16} has
its natural expression in coordinate representation \cite{13}, i.e.,
\begin{equation}
S_{1}\left( \mu \right) =\frac{1}{\sqrt{\mu }}\int_{-\infty }^{\infty }%
\mathtt{d}x\left\vert \frac{x}{\mu }\right\rangle \left\langle x\right\vert ,
\label{f12}
\end{equation}%
which leads to $\left\vert \nu y\right\rangle =\frac{1}{\sqrt{\nu }}%
S_{1}^{-1}\left( \nu \right) \left\vert y\right\rangle ,$ $\left\langle \mu
x\right\vert =\frac{1}{\sqrt{\mu }}\left\langle x\right\vert S_{1}\left( \mu
\right) ,$ so Eq.(\ref{f11}) can be decomposed into
\begin{equation}
\mathcal{H}_{\alpha }\left( \mu ,\nu \right) =S_{1}^{-1}\left( \nu \right)
\mathcal{F}_{\alpha }S_{1}\left( \mu \right) =S_{1}^{-1}\left( \nu \right)
e^{\mathtt{i}\alpha a^{\dagger }a}S_{1}\left( \mu \right) ,  \label{f14}
\end{equation}%
where $\mathcal{F}_{\alpha }$ is given by%
\begin{equation}
\mathcal{F}_{\alpha }\equiv \sqrt{\frac{e^{\mathtt{i}\left( \frac{\pi }{2}%
-\alpha \right) }}{2\pi \sin \alpha }}\int_{-\infty }^{\infty }e^{-\frac{%
\mathtt{i}(x^{2}+y^{2})}{2\tan \alpha }+\frac{\mathtt{i}xy}{\sin \alpha }%
}\left\vert y\right\rangle \left\langle x\right\vert \mathtt{d}x\mathtt{d}%
y=e^{\mathtt{i}\alpha a^{\dagger }a},  \label{f15}
\end{equation}%
this integral result can be obtained by using a similar way to deriving Eq.(%
\ref{f6}). Thus we see that the CFrHO can be decomposed as a fractional
operator and two-single-mode squeezing operators.

Using the decomposition of the CFrHO in Eq.(\ref{f14}), and noticing that $%
S_{1}\left( \mu \right) XS_{1}^{-1}\left( \mu \right) =\mu X,$ $S_{1}\left(
\mu \right) PS_{1}^{-1}\left( \mu \right) =P/\mu ,$ and $e^{\mathtt{i}\alpha
a^{\dagger }a}ae^{-\mathtt{i}\alpha a^{\dagger }a}=ae^{-\mathtt{i}\alpha },$
which leads to
\begin{equation}
e^{\mathtt{i}\alpha a^{\dagger }a}Xe^{-\mathtt{i}\alpha a^{\dagger
}a}=\allowbreak X\cos \alpha +P\sin \alpha ,\text{ }e^{\mathtt{i}\alpha
a^{\dagger }a}Pe^{-\mathtt{i}\alpha a^{\dagger }a}=\allowbreak P\cos \alpha
-X\sin \alpha ,  \label{f15a}
\end{equation}
thus we have%
\begin{eqnarray}
\mathcal{H}_{\alpha }\left( \mu ,\nu \right) X\mathcal{H}_{\alpha }^{\dagger
}\left( \mu ,\nu \right) &=&\mu S_{1}^{-1}\left( \nu \right) \left( X\cos
\alpha +P\sin \alpha \right) S_{1}\left( \nu \right)  \notag \\
&=&\frac{\mu }{\nu }X\cos \alpha +\mu \nu P\sin \alpha ,  \label{f16} \\
\mathcal{H}_{\alpha }\left( \mu ,\nu \right) P\mathcal{H}_{\alpha }^{\dagger
}\left( \mu ,\nu \right) &=&\frac{\allowbreak \nu }{\mu }P\cos \alpha -\frac{%
X}{\mu \nu }\sin \alpha ,  \label{f17}
\end{eqnarray}%
from which we see that the CFrHO plays the role of combining the coordinate
operator $X$ and momentum operator $P$ in a certain way (\ref{f16})-(\ref%
{f17}), i.e., including the squeezing and the rotation. In paticular, when $%
\alpha =\frac{\pi }{2},$ Eqs.(\ref{f16})-(\ref{f17}) become%
\begin{equation}
\mathcal{H}_{\frac{\pi }{2}}\left( \mu ,\nu \right) X\mathcal{H}_{\frac{\pi
}{2}}^{\dagger }\left( \mu ,\nu \right) =\mu \nu P,\text{ }\mathcal{H}_{%
\frac{\pi }{2}}\left( \mu ,\nu \right) P\mathcal{H}_{\frac{\pi }{2}%
}^{\dagger }\left( \mu ,\nu \right) =-\frac{X}{\mu \nu },  \label{f18}
\end{equation}%
i.e., the mutual exchanging of coordinate-momentum operators.

On the other hand, there is a most important feature of the FrFT is that the
FrFT obeys the additivity rule, i.e., two successive FrFT of order $\alpha $%
\ and $\beta $ makes up the FrFT of order $\alpha +\beta $. Then a question
naturally arises: Is the two successive CFrHOs still a CFrHO? To answer this
question, we examine the direct product $\mathcal{H}_{\alpha }\left( \mu
,\nu \right) \otimes \mathcal{H}_{\beta }\left( \mu ^{\prime },\nu ^{\prime
}\right) $. Using Eq.(\ref{f14}) it is easily seen that when $\mu =$ $\nu
^{\prime }$ there is an additivity of operator as follows%
\begin{eqnarray}
\mathcal{H}_{\alpha }\left( \mu ,\nu \right) \otimes \mathcal{H}_{\beta
}\left( \mu ^{\prime },\mu \right) &=&S_{1}^{-1}\left( \nu \right) e^{%
\mathtt{i}\alpha a^{\dagger }a}S_{1}\left( \mu \right) S_{1}^{-1}\left( \mu
\right) e^{\mathtt{i}\beta a^{\dagger }a}S_{1}\left( \mu ^{\prime }\right)
\notag \\
&=&\mathcal{H}_{\alpha +\beta }\left( \mu ^{\prime },\nu \right) ,
\label{f19}
\end{eqnarray}%
which can be seen as the additivity property of the CFrHOs. Here it should
be pointed out that the condition of additivitive operator for the CFrHOs is
that the parameter $\mu $ of the prior cascade operator should be equal to
the parameter $\nu ^{\prime }$ of the next one, i.e., $\mu =\nu ^{\prime }.$
This can be clearly seen from the viewpoint of classical optics transform.

\section{Two-mode CFrHT}

Next, we shall extend the single-mode CFrHT to two-mode case by using the
entangled state representation \cite{17},%
\begin{equation}
\left\vert \eta \right\rangle =\exp \left\{ -\frac{1}{2}\left\vert \eta
\right\vert ^{2}+\eta a_{1}^{\dagger }-\eta ^{\ast }a_{2}^{\dagger
}+a_{1}^{\dagger }a_{2}^{\dagger }\right\} \left\vert 00\right\rangle ,
\label{f20}
\end{equation}%
where $\left\vert \eta =\eta _{1}+\mathtt{i}\eta _{2}\right\rangle $ is the
common eigenvector of two-particle's relative coordinate $X_{1}-X_{2}$ and
total momentum $P_{1}+P_{2},$%
\begin{equation}
\left( X_{1}-X_{2}\right) \left\vert \eta \right\rangle =\sqrt{2}\eta
_{1}\left\vert \eta \right\rangle ,\left( P_{1}+P_{2}\right) \left\vert \eta
\right\rangle =\sqrt{2}\eta _{2}\left\vert \eta \right\rangle ,  \label{f21}
\end{equation}%
and $\left\vert \eta \right\rangle $ possesses the completeness and the
orthogonality,
\begin{equation}
\int_{-\infty }^{\infty }\frac{\mathtt{d}^{2}\eta }{\pi }\left\vert \eta
\right\rangle \left\langle \eta \right\vert =1,\text{ }\left\langle \eta
\right\vert \left. \eta ^{\prime }\right\rangle =\pi \delta \left( \eta
-\eta ^{\prime }\right) \delta \left( \eta ^{\ast }-\eta ^{\prime \ast
}\right) .  \label{f22}
\end{equation}%
In a similar way to introducing Eq.(\ref{f3}), we examine the following
transform,%
\begin{equation}
\mathcal{H}_{\alpha }^{C}\left( \mu ,\nu \right) \left\vert \eta
\right\rangle =\frac{e^{\mathtt{i}\left( \frac{\pi }{2}-\alpha \right) }}{%
2\mu \nu \sin \alpha }\int \frac{\mathtt{d}^{2}\eta ^{\prime }}{\pi }e^{-%
\frac{\mathtt{i}\left( \allowbreak \left\vert \eta ^{\prime }\right\vert
^{2}/\nu ^{2}+\left\vert \eta \right\vert ^{2}/\mu ^{2}\allowbreak \right) }{%
2\tan \alpha }+\frac{\mathtt{i}\left( \eta ^{\prime }{}^{\ast }\allowbreak
\eta +\eta ^{\ast }\allowbreak \eta ^{\prime }\right) }{2\mu \nu \sin \alpha
}}\left\vert \eta ^{\prime }\right\rangle .  \label{f23}
\end{equation}%
Using Eq.(\ref{f22}), one can see that $\mathcal{H}_{\alpha }^{C}\left( \mu
,\nu \right) $ is a unitary operator, i.e., $\mathcal{H}_{\alpha }^{C}\left[
\mathcal{H}_{\alpha }^{C}\right] ^{\dag }=\left[ \mathcal{H}_{\alpha }^{C}%
\right] ^{\dag }\mathcal{H}_{\alpha }^{C}=1$. Here we should emphasize that,
the exponential item in the right hand side of Eq.(\ref{f23}) can be
decomposed into a direct product of two exponential items in the right hand
side of Eq.(\ref{f3}), but $\left\vert \eta \right\rangle $ is an entangled
state (not the direct product of two single-mode coordinate states, which
can be seen clearly from its Schmidt decomposition \cite{18}), thus this is
a nontrivial extension from single-mode case to two-mode case.

Performing a similar procedure to single-mode case, i.e., noticing that the
two-mode squeezing operator has its natural expression in the entangled
state representation,%
\begin{equation}
S_{2}\left( \mu \right) =\exp \left[ \left( a_{1}^{\dagger }a_{2}^{\dagger
}-a_{1}a_{2}\right) \ln \mu \right] =\frac{1}{\mu }\int \frac{d^{2}\eta }{%
\pi }\left\vert \frac{\eta }{\mu }\right\rangle \left\langle \eta
\right\vert ,  \label{f24}
\end{equation}%
which leads to $\frac{1}{\nu }\left\vert \frac{\eta ^{\prime }}{\nu }%
\right\rangle =S_{2}\left( \nu \right) \left\vert \eta ^{\prime
}\right\rangle ,\frac{1}{\mu }\left\vert \frac{\eta }{\mu }\right\rangle
=S_{2}\left( \mu \right) \left\vert \eta \right\rangle ,$ then\ using the
completeness of $\left\vert \eta \right\rangle $\ and $\left\vert
00\right\rangle \left\langle 00\right\vert =\colon e^{-a^{\dag }a-b^{\dag
}b}\colon $ and the orthogonality in Eq.(\ref{f22}) we can further decompose
the operator $\mathcal{H}_{\alpha }^{C}\left( \mu ,\nu \right) $ into the
following form,%
\begin{equation}
\mathcal{H}_{\alpha }^{C}\left( \mu ,\nu \right) =S_{2}^{\dagger }\left( \nu
\right) \mathcal{F}_{\alpha }^{C}S_{2}\left( \mu \right) =S_{2}^{\dagger
}\left( \nu \right) \exp \left\{ \mathtt{i}\alpha \left( a_{1}^{\dagger
}a_{1}+a_{2}^{\dagger }a_{2}\right) \right\} S_{2}\left( \mu \right) ,
\label{f25}
\end{equation}%
where the operatror $\mathcal{F}_{\alpha }^{C}$ is given by%
\begin{eqnarray}
\mathcal{F}_{\alpha }^{C} &=&\frac{e^{\mathtt{i}\left( \frac{\pi }{2}-\alpha
\right) }}{2\sin \alpha }\int \frac{\mathtt{d}^{2}\eta ^{\prime }\mathtt{d}%
^{2}\eta }{\mu ^{2}\nu ^{2}\pi }e^{-\frac{\mathtt{i}\left( \allowbreak
\left\vert \eta ^{\prime }\right\vert ^{2}/\nu ^{2}+\left\vert \eta
\right\vert ^{2}/\mu ^{2}\allowbreak \right) }{2\tan \alpha }+\frac{\mathtt{i%
}\left( \eta ^{\prime }{}^{\ast }\allowbreak \eta +\eta ^{\ast }\allowbreak
\eta ^{\prime }\right) }{2\mu \nu \sin \alpha }}\left\vert \frac{\eta
^{\prime }}{\nu }\right\rangle \left\langle \frac{\eta }{\mu }\right\vert
\notag \\
&=&\exp \left\{ \mathtt{i}\alpha \left( a_{1}^{\dagger }a_{1}+a_{2}^{\dagger
}a_{2}\right) \right\} .  \label{f26}
\end{eqnarray}%
Thus we see that the two-mode CFrHO can be decomposed into the form in Eq.(%
\ref{f25}), i.e., two fractional operators and two two-mode squeezing
operators.

This is a convient expression for further deriving the transforms and the
condition of additivitive operator. In fact, using Eqs.(\ref{f25}), (\ref%
{f15a}) and Eqs.(\ref{f21}), (\ref{f22}) leading to%
\begin{eqnarray}
S_{2}\left( \mu \right) \left( X_{1}-X_{2}\right) S_{2}^{\dagger }\left( \mu
\right) &=&\mu \left( X_{1}-X_{2}\right) ,\text{ }S_{2}\left( \mu \right)
\left( P_{1}+P_{2}\right) S_{2}^{\dagger }\left( \mu \right) =\mu \left(
P_{1}+P_{2}\right) ,  \label{f27} \\
S_{2}\left( \mu \right) \left( X_{1}+X_{2}\right) S_{2}^{\dagger }\left( \mu
\right) &=&\frac{1}{\mu }\left( X_{1}+X_{2}\right) ,\text{ }S_{2}\left( \mu
\right) \left( P_{1}-P_{2}\right) S_{2}^{\dagger }\left( \mu \right) =\frac{1%
}{\mu }\left( P_{1}-P_{2}\right) ,  \label{f28}
\end{eqnarray}
we have%
\begin{eqnarray}
\mathcal{H}_{\alpha }^{C}\left( X_{1}-X_{2}\right) \left[ \mathcal{H}%
_{\alpha }^{C}\right] ^{\dag } &=&\mu S_{2}^{\dagger }\left( \nu \right)
\mathcal{F}_{\alpha }^{C}\left( X_{1}-X_{2}\right) \left[ \mathcal{F}%
_{\alpha }^{C}\right] ^{\dag }S_{2}\left( \nu \right)  \notag \\
&=&\mu S_{2}^{\dagger }\left( \nu \right) \left( \left( X_{1}-X_{2}\right)
\cos \alpha +\left( P_{1}-P_{2}\right) \sin \alpha \right) S_{2}\left( \nu
\right)  \notag \\
&=&\frac{\mu }{\nu }\left( X_{1}-X_{2}\right) \cos \alpha +\mu \nu \left(
P_{1}-P_{2}\right) \sin \alpha ,  \label{f29} \\
\mathcal{H}_{\alpha }^{C}\left( X_{1}+X_{2}\right) \left[ \mathcal{H}%
_{\alpha }^{C}\right] ^{\dag } &=&\frac{\nu }{\mu }\left( X_{1}+X_{2}\right)
\cos \alpha +\frac{1}{\mu \nu }\left( P_{1}+P_{2}\right) \sin \alpha ,
\label{f30}
\end{eqnarray}%
and
\begin{eqnarray}
\mathcal{H}_{\alpha }^{C}\left( P_{1}-P_{2}\right) \left[ \mathcal{H}%
_{\alpha }^{C}\right] ^{\dag } &=&\frac{\nu }{\mu }\left( P_{1}-P_{2}\right)
\cos \alpha -\frac{1}{\mu \nu }\left( X_{1}-X_{2}\right) \sin \alpha ,
\label{f31} \\
\mathcal{H}_{\alpha }^{C}\left( P_{1}+P_{2}\right) \left[ \mathcal{H}%
_{\alpha }^{C}\right] ^{\dag } &=&\frac{\mu }{\nu }\left( P_{1}+P_{2}\right)
\cos \alpha -\mu \nu \left( X_{1}+X_{2}\right) \sin \alpha .  \label{f32}
\end{eqnarray}%
From Eqs.(\ref{f29})-(\ref{f32}) it is easy to see that when $\alpha =\pi
/2, $ the role of $\mathcal{H}_{\pi /2}^{C}$ is just exchanging $\left(
X_{1}-X_{2}\right) $ and $\left( P_{1}-P_{2}\right) $, $\left(
X_{1}+X_{2}\right) $ and $\left( P_{1}+P_{2}\right) ;$ while for $\alpha
=\pi ,$ $\mathcal{H}_{\pi }^{C}$ can be seen as an identity operator.

In addition, from the decomposition (\ref{f25}) one can see that the direct
product $\mathcal{H}_{\alpha }^{C}\left( \mu ,\nu \right) \otimes \mathcal{H}%
_{\beta }^{C}\left( \mu ^{\prime },\nu ^{\prime }\right) $ satisfies the
additivity rule when $\mu =$ $\nu ^{\prime }$, i.e.,
\begin{equation}
\mathcal{H}_{\alpha }^{C}\left( \mu ,\nu \right) \otimes \mathcal{H}_{\beta
}^{C}\left( \mu ^{\prime },\nu ^{\prime }\right) =\mathcal{H}_{\alpha +\beta
}^{C}\left( \mu ^{\prime },\nu \right) .  \label{f33}
\end{equation}

\section{Measurements for the output states from the CFrHT}

The measurement for quantum state plays an important role in quantum
computation and quantum imfromation. When a quantum state $\left\vert
f\right\rangle $ is transformed by the CFrHO, then what is the measurement
result with continuous orthogonal basis? For single-mode case, the output
state from the CFrHT is $\left\vert g\right\rangle _{out}=\mathcal{H}%
_{\alpha }\left( \mu ,\nu \right) \left\vert f\right\rangle $. The
measurement basis is choosen as a coordiante eigenvector, then the
measurement result is given by%
\begin{eqnarray}
\left\langle x\right. \left\vert g\right\rangle _{out} &=&\left\langle
x\right\vert \mathcal{H}_{\alpha }\left( \mu ,\nu \right) \left\vert
f\right\rangle  \notag \\
&=&\int_{-\infty }^{\infty }\mathtt{d}x^{\prime }\left\langle x\right\vert
\mathcal{H}_{\alpha }\left( \mu ,\nu \right) \left\vert x^{\prime
}\right\rangle \left\langle x^{\prime }\right. \left\vert f\right\rangle
\notag \\
&=&\sqrt{\frac{e^{\mathtt{i}\left( \frac{\pi }{2}-\alpha \right) }}{2\pi \mu
\nu \sin \alpha }}\int_{-\infty }^{\infty }f\left( x^{\prime }\right) e^{-%
\frac{\mathtt{i}\left( x^{\prime 2}/\mu ^{2}+x^{2}/\nu ^{2}\right) }{2\tan
\alpha }+\frac{\mathtt{i}x^{\prime }x}{\mu \nu \sin \alpha }}\mathtt{d}%
x^{\prime },  \label{f34}
\end{eqnarray}%
which just corresponds to a generalized fractional Fourier transform of wave
function $f\left( x^{\prime }\right) =\left\langle x^{\prime }\right.
\left\vert f\right\rangle .$

For two-mode case, the measurement result by two-mode entangled state Bell
basis is%
\begin{eqnarray}
\left\langle \eta ^{\prime }\right. \left\vert g\right\rangle _{out}
&=&\left\langle \eta ^{\prime }\right\vert \mathcal{H}_{\alpha }^{C}\left(
\mu ,\nu \right) \left\vert f\right\rangle  \notag \\
&=&\int_{-\infty }^{\infty }\frac{\mathtt{d}^{2}\eta }{\pi }\left\langle
\eta ^{\prime }\right\vert \mathcal{H}_{\alpha }^{C}\left( \mu ,\nu \right)
\left\vert \eta \right\rangle \left\langle \eta \right. \left\vert
f\right\rangle  \notag \\
&=&\frac{e^{\mathtt{i}\left( \frac{\pi }{2}-\alpha \right) }}{2\mu \nu \sin
\alpha }\int_{-\infty }^{\infty }\frac{\mathtt{d}^{2}\eta }{\pi }e^{-\frac{%
\mathtt{i}\left( \allowbreak \left\vert \eta ^{\prime }\right\vert ^{2}/\nu
^{2}+\left\vert \eta \right\vert ^{2}/\mu ^{2}\allowbreak \right) }{2\tan
\alpha }+\frac{\mathtt{i}\left( \eta ^{\prime }{}^{\ast }\allowbreak \eta
+\eta ^{\ast }\allowbreak \eta ^{\prime }\right) }{2\mu \nu \sin \alpha }%
}f\left( \eta \right) ,  \label{f35}
\end{eqnarray}%
which is just a generalized complex fractional Fourier transform, and the
wave function $f\left( \eta \right) $ is the projection of quantum state $%
\left\vert f\right\rangle $ on $\left\langle \eta \right\vert $. From Eqs.(%
\ref{f34}) and (\ref{f35}) we can clearly see that the generalized FrFT of
the wavefunction for any quantum state $\left\vert f\right\rangle $ in
coordinate/entangled state corresponds to the wavefunction of
Hadamard-transformed ($\mathcal{H}_{\alpha }\left( \mu ,\nu \right)
\left\vert f\right\rangle $) in coordinate/entangled state. In other words,
the generalized FrFT of wavefunction is just the matrix element of CFrHO in $%
\left\langle x\right\vert $ ($\left\langle \eta ^{\prime }\right\vert $) and
$\left\vert f\right\rangle $.

\section{Conclusion}

Based on quantum Hadamard transform, we have introduced the quantum
fractional Hadamard transform with continuous variables. It is found that
the corresponding quantum fractional Hadamard operator can be decomposed
into a single-mode fractional operator $e^{\mathtt{i}\alpha a^{\dagger }a}$
and two single-mode squeezing operators. The two-mode fractional Hadamard
transform is also introduced by using the bipartite entangled state
representation. It is shown that the corresponding transform operatror
involves two single-mode fractional operators and two two-mode squeezing
operators. For any quantum state vector $\left\vert f\right\rangle $, the
measurement results for the transformed quantum state (for instance $%
\left\vert g\right\rangle _{out}=\mathcal{H}_{\alpha }\left( \mu ,\nu
\right) \left\vert f\right\rangle )$ by continuous coordinate state $%
\left\vert x\right\rangle $ (or bipartite entangled state $\left\vert \eta
\right\rangle )$ just corresponds to a generalized (complex) fractional
Fourier transform. In addition, the new transformation gives us more
flexibility to represent signals in the fractional Hadamard domain with
extra freedom provided by an angle, and two-squeezing parameters. For more
discussions about the optical transforms in the context of quantum optics
and the discrete fractional Hadmard transform, we refer to Refs.\cite%
{19,20,21}.

\end{document}